\documentclass[a4paper,twocolumn,aps,prb,floatfix,showpacs,preprintnumbers]{revtex4}
\usepackage{amsmath}
\usepackage{amssymb}
\usepackage{graphicx}
\usepackage{color}

\begin{document}

\title{Electron interaction, charging and screening in grain boundaries in
graphene}
\author{S. Ihnatsenka }
\email{sihnatse@sfu.ca}
\affiliation{Department of Physics, Simon Fraser University, Burnaby, British
Columbia, Canada V5A 1S6}
\author{I. V. Zozoulenko}
\email{igor.zozoulenko@liu.se}
\affiliation{Organic Electronics, Department of Science and Technology (ITN),
Link\"{o}ping University, 60174 Norrk\"{o}ping, Sweden}
\date{\today}

\begin{abstract}
Electronic, transport, and spin properties of grain boundaries (GBs) are
investigated in electrostatically doped graphene at finite electron
densities within the Hartree and Hubbard approximations. We demonstrate that
depending on the character of the GBs, the states residing on them can have a
metallic character with a zero group velocity or can be fully populated
losing the ability to carry a current. These states show qualitatively
different features in charge accumulation and spin polarization. We also
demonstrate that the semiclassical Thomas-Fermi approach provides a
satisfactory approximation to the calculated self-consistent potential. The
conductance of GBs is reduced due to enhanced backscattering from this
potential.
\end{abstract}
\pacs{81.05.ue, 73.22.Pr, 72.80.Vp, 72.10.Fk, 73.20.Hb}

\maketitle

\section{Introduction}
During recent years chemical vapor deposition (CVD)
on transition metals has emerged as one of the most attractive methods for
scalable graphene production\cite{Li,Kim}. The advantages of this method is
in its low cost, the possibility to grow large graphene sheets (tens of inches),
and the ease of its transfer onto other substrates. Due to features of the
growth process, CVD-grown graphene is polycrystalline, consisting of grains of
various crystal orientations separated by one-dimensional extended line
defects representing grain boundaries (GBs)\cite{Huang,KimCrommie}.
Significant evidence has accumulated by now that the GBs strongly affect
electrical transport\cite{YazyevNature,Tapaszto,Ahmad,Yu,Tsen,Koepke} and
represent the limiting scattering mechanism of the electronic mobility in
CVD-grown graphene\cite{Ferreira,Peres,Tuan2013,Taras_Aires}. This provides
a strong motivation for investigation of morphological, electronic and spin
properties of GBs. A number of studies have been recently reported
addressing the band structure\cite{Bahamon,Alexandre,Botello,Song,Liwei},
spin polarization\cite{Botello,Alexandre,Okada,Kou}, electron transport and
scattering\cite{Ferreira,Peres,Liwei} in GBs. However, all these studies
were limited to the case of electrically neutral graphene, and very little
is presently known on how the electronic and transport properties of GBs are
modified at nonzero electron densities (i.e., away from the Dirac point). At
the same time, the effect of a finite electron density is of the utmost
importance for the understanding of electron scattering by the GBs. Indeed, due to filling of quasibound states residing on them by
electrons from the bulk, the GBs transform into charged lines which are
believed to be responsible for the impediment of electron transport in CDV-grown
graphene. Note that a local self-doping of individual GBs (i.e. transfer of
electrons from the bulk to the states at GBs in a nominally neutral sample)
has been recently observed by means of STM measurements\cite{Tapaszto}; it
has also been argued that by doping by electrons from the bulk, the GBs can
act as quasi-dimensional metallic wires\cite{Lahiri,Liwei,Peres}. It is
noteworthy that quasi-one-dimensional localized states of a related nature
can reside on domain walls\cite{Semenoff}, graphene nanoroads\cite{Jung},
and p-n junctions in bilayer graphene\cite{Qiao}.

In the present work we depart from a conventional model of neutral graphene
at the half filling and investigate how electronic and transport properties
of GBs are affected by the presence of interacting electrons. We discuss how
the charging of GBs evolves with the electron density and compare our findings
with a semiclassical Thomas-Fermi (TF) model of screening. We also demonstrate
that charging at finite electron densities leads to qualitatively new
features in the band structure, and transport properties of grain boundary
are strongly modified in a comparison to the noninteracting description.

\section{Basics}
\begin{figure}[th]
\includegraphics[width=\columnwidth]{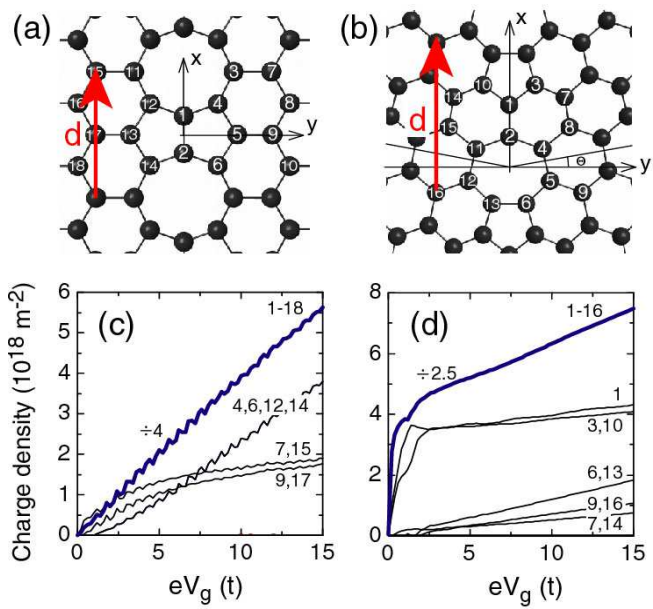}
\caption{(color online)  Atomic geometries of (2,0) GB and (2,1) GB (panels (a) and (b) respectively); the
arrows correspond to the repeated vectors (2,0) and (2,1). Thick blue lines in (c) and (d) show the total occupancies for (2,0) and (2,1) GBs as a function of $E_{F}=eV_{g}$ (divided respectively by factors of 4 and 2.5). Thin black lines show occupancies of individual carbon atoms (enumerated in (a) and (b)). }
\label{fig:geometry}
\end{figure}

In graphene, GBs represent one dimensional dislocations
defined by interfaces between two domains of material with different
crystallographic orientations. The later are characterized by the angles $%
\theta _{L}$ and $\theta _{R}$ between the corresponding crystallographic
directions in two domains and the normal to the boundary line, Fig. \ref{fig:geometry}. (See also an illustration in Fig. \ref{fig:SI:1} in the Appendix). The periodicity of the dislocation is defined by
the translation vectors $(n,m)$ of the length $d$ belonging to the
crystalline domains and oriented along the boundary line. In the present
study we consider two representative GBs, (2,0) and (2,1), shown respectively in Fig. \ref{fig:geometry} (a) and (b). The first one,
(2,0), consists of domains with the aligned crystallographic orientations $%
\theta _{L}=\theta _{R}=\theta =0^{\circ }$ ($d\approx $ 0.5nm) and
separated by a zigzag-oriented interface of one octagon and two side-sharing
pentagons.\cite{Botello,Alexandre,Song,Bahamon,Lahiri} The repeat vector
(2,1) of the second one implies $\theta _{L}=\theta _{R}=\theta =10.9^{\circ }$ (%
$d\approx $ 0.65nm) and its interface region includes pentagon-heptagon
pairs.\cite{Alexandre,YazyevNature} We would like to note that while we
study two representative GBs, (2,0) and (2,1) (corresponding to aligned and
misaligned crystallographic orientations), we believe that our findings are
generic and remain valid for other GBs in graphene.

\begin{figure}[th]
\includegraphics[width=\columnwidth]{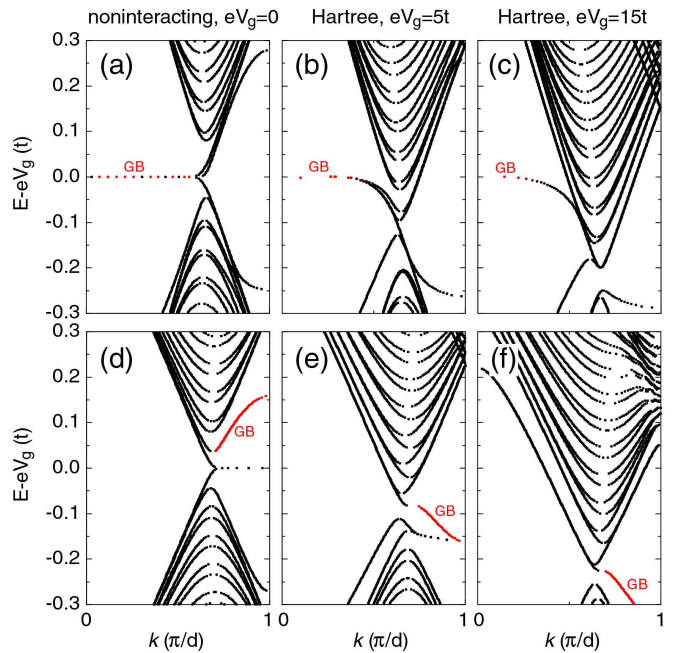}
\caption{(color online) Evolution of the band structure of (a)-(c) the
(2,0) GB state and (d)-(f) the (2,1) GB state upon change of the gate voltage $V_{g}$
calculated in the Hartree approximation using Hamiltonian Eq. (\ref{eq:Hamiltonian}). States residing at GBs are drawn in red and marked by \textquotedblleft
GB\textquotedblright.}
\label{fig:band}
\end{figure}

For the case of spinless electrons we use a standard $p$-orbital
tight-binding Hamiltonian\cite{Shylau_Cap},
\begin{equation}
H_{0}=\sum_{\mathbf{r}}V_{\mathbf{r}}^{Hartree}a_{\mathbf{r}}^{+}a_{\mathbf{r%
}}-\sum_{\mathbf{r,r+\Delta }}t_{\mathbf{r,r+\Delta }}a_{\mathbf{r}}^{+}a_{%
\mathbf{r}+\Delta },  \label{eq:Hamiltonian}
\end{equation}%
where $V_{\mathbf{r}}^{Hartree}$ is the Hartree potential at the site $%
\mathbf{r}$, and $t_{\mathbf{r,r+\Delta }}$is the overlap integral between the
neighboring sites $\mathbf{r}$ and $\mathbf{r+\Delta }.$ The Hartree
potential results from the Coulomb interaction between extra charges in the
system,
\begin{equation}
V_{\mathbf{r}}^{Hartree}=\frac{e^{2}}{4\pi \varepsilon _{0}\varepsilon _{r}}%
\sum_{\mathbf{r}^{^{\prime }}\neq \mathbf{r}}n_{\mathbf{r}^{^{\prime
}}}\left( \frac{1}{|\mathbf{r}-\mathbf{r}^{^{\prime }}|}-\frac{1}{\sqrt{|%
\mathbf{r}-\mathbf{r}^{^{\prime }}|^{2}+4b^{2}}}\right) ,
\label{eq:V_Hartree}
\end{equation}%
where $n_{\mathbf{r}^{^{\prime }}}$\ is the local electron occupation, and
the second term in the parentheses corresponds to the mirror charges \cite%
{Shylau_Cap}. (In our calculation we assume that a graphene sheet is
separated from a back gate by a dielectric of width $b=50$ nm with the
relative permittivity $\varepsilon _{r}=3.9$). The summation in the Hartree
potential, Eq. (\ref{eq:V_Hartree}), runs over the entire ribbon. In order
to calculate $t_{\mathbf{r,r+\Delta }}$ for the GBs studied in this paper we
performed \textit{ab initio} geometry relaxations based on the density
functional theory using the Gaussian 09 software package.\cite{Gaussian}
Away of the GB, $t_{\mathbf{r,r+\Delta }}=t=2.7$ eV. Details of
computations and the calculated values of the transfer integrals are
presented in the Appendix. The number of
excess electrons at site $\mathbf{r}$ reads, $n_{\mathbf{r}%
}=\int_{-\infty }^{\infty }\rho (\mathbf{r},E)f_{FD}(E,E_{F})dE-n_{\text{ions%
}},$ where $\rho (\mathbf{r},E)$ is the energy dependent local density of
states, $f_{FD}(E,E_{F})$ is the Fermi-Dirac distribution function, $%
E_{F}=eV_{g}$ is the Fermi energy the value of which is adjusted by the gate
voltage $V_{g}$, and $n_{\text{ions}}=3.8\times 10^{19}$m$^{-2}$ is the
positive charge background of ions. The Bloch states, the electron densities, and
the band structure are calculated self-consistently using the Green's
function technique as described in Refs. \onlinecite{Shylau_Cap,Ihnatsenka2012}. The
conductance calculations with a self-consistent potential are performed on
the basis of the Landauer formalism using the standard recursive Greens function
technique as described in Ref. \cite{conductance}. The band structure
calculations are performed in the ribbon geometry with the GB residing in
the middle of the ribbon which is infinite in the $x$-direction and has a
finite width of $20$ nm in the transverse $y$-direction.

For the case of electrons with spin ($\sigma =\uparrow ,\downarrow $) we introduce spin-dependent
electron densities $n_{\mathbf{r}}^{\sigma }$ and use the same formalism as
described above with a Hubbard Hamiltonian of the form $H=H^{\uparrow
}+H^{\downarrow }$, \cite{Wehling,Ihnatsenka2012}

\begin{equation}
H^{\sigma}=H_{0}+V_{Hubb}^{\sigma }; \ V_{Hubb}^{\sigma }=Un_{\mathbf{r}}^{\sigma ^{\prime}},
\label{eq:Hubbard}
\end{equation}%
where $H_{0}$ is given by Eq. (\ref%
{eq:Hamiltonian}) and  the Hubbard constant $U=t$.\cite{Yazyev}

\section{Results and discussion.}
We start with the case of spinless
electrons described by the Hamiltonian (\ref{eq:Hamiltonian}). Figures %
\ref{fig:band}(a)-(c) show the band structures of the (2,0) GB for different
gate voltages $V_{g}$. For $V_{g}=0$ (Fig. \ref{fig:band} (a)) the system
remains neutral and the results of the Hartree approach corresponds to the case of non-interacting electrons. The flat band at $E-E_{F}=0$ is nearly degenerate and
corresponds to three states, one residing at the GB and two at the zigzag
edges of the ribbon. (Note that even though the zigzag edge states
contribute to the band diagram, we verified, by choosing a larger
computational domain, that their overall effect on electrostatics and
electronic properties of the state at the GB is negligibly small). An
inspection of the wavefunctions shows that the flat band of the GB is
associated with an exponentially localized Bloch state which originates from
the zigzag topology of the interface similar to the zigzag edges of zigzag
ribbons. As $V_{g}$ increases the band structure changes substantially, see
Fig. \ref{fig:band} (b),(c). The most distinct feature of the band diagram is
that the state at the GB gets pinned to $E_{F}$ and remains partially filled
at any $V_{g}$. This is because this state is flat and therefore has a high
density of states (DOS). As a result, electrons filling this state can
easily screen the external potential, which results in metallic behavior and
pinning.

The state residing at the (2,1) GB also shows exponential localization.
However, features and evolution of the corresponding band in the dispersion
relation are different from those of the (2,0) state, see the state marked
``GB'' in Fig. \ref{fig:band}(d)-(f). At $V_{g}=0$ the state residing at the GB
is practically empty as it lies above $E_{F}$, Fig. \ref{fig:band}(d). In
contrast to the (2,0) state, this state does not have a metallic character
with a high DOS, and therefore it can not screen the applied potential.
Hence, with application of $V_{g}$ the (2,1) GB state gets quickly
populated, and a corresponding dispersion curve bends down and moves below $%
E_{F}$, see Fig. \ref{fig:band}(e),(f). Note that a flat band at $E_{F}=0$ in
Fig. \ref{fig:band}(d) corresponds to the edge states of the zigzag nanoribbon.

\begin{figure}[th]
\includegraphics[width=0.5\textwidth]{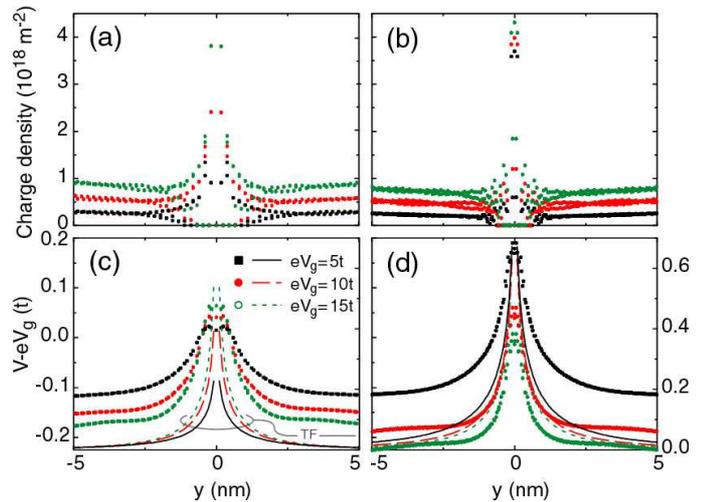}
\caption{(color online) (a)-(b) The charge density and (c)-(d) the potential
in a cross-section of the ribbon calculated in the Hartree approximation for
respectively (2,0) and (2,1) GBs for different applied gate voltages $%
eV_{g}=5,10,15t$. Thin solid, dashed and dotted lines in (c) and (d)
illustrate the results of the Thomas-Fermi approximation (Eq. \protect\ref%
{eq:V_TF}). Extracted linear
charge densities at GBs and the bulk electron densities (i.e. away from the
GBs) are $\protect\lambda/a=1.3,2.6,3.9\times10^{18}$ m$^{-2}$ and $n$=3, 6,
9$\times10^{17}$ m$^{-2}$ for (c) and $\protect\lambda/a$=3.7, 4, 4.3$%
\times10^{18}$ m$^{-2}$ and $n$=2.5, 5, 7.5$\times10^{17}$ m$^{-2}$ for (d).
Energy scale for the TF potentials (in units of $t$) is on the right of (c) and (d).}
\label{fig:charge}
\end{figure}

Let us now discuss charge accumulation at the GBs due to the filling of the
quasibound states residing on them. Figure \ref{fig:geometry} shows that the
charge density at the (2,0) GB grows linearly with an increase of the gate
voltage $V_{g}$. In contrast, the charge density at the (2.1) GB stays
practically constant showing only a slow increase following the increase of
the overall electron density in the ribbon as $V_{g}$ grows. This difference
in the charge accumulation can be traced to different behavior of the band
dispersions of the states residing at the (2,0) and (2,1) GBs discussed above.
The linear charge accumulation on the (2,0) GB occurs because the
corresponding state has a metallic character with a high DOS and therefore
it remains only partially filled. As a result, an increase of the gate
voltage leads to a gradual population of this state. On the contrary, with
application of the gate voltage the state residing at the (2,1) GB becomes
practically fully populated and immediately moves below $E_{F}$. Hence, a
further increase of $V_{g}$ has very little effect on the charge accumulated
at the (2,1) GB. It should be noted that for both types of GBs considered
here the local density of states (LDOS) and therefore the accumulated charge
strongly depend on a site position, see Fig. \ref{fig:geometry}
(c),(d).

As mentioned in the Introduction, scattering at charged line defects is regarded
as the limiting factor for the mobility in CVD graphene. An expression for
the scattering potential can be obtained within the semiclassical
TF approximation describing the screening of an extended charged
line defect by the surrounding electron gas \cite{Ferreira,Taras_Aires},
\begin{align}
V_{TF}(x)=& \frac{\lambda e}{2\pi \varepsilon _{0}\varepsilon _{\text{r}}}%
\Bigl\{-\cos \left( q_{\text{TF}}x\right) \text{Ci}\left( q_{\text{TF}%
}x\right) +  \notag \\
& +\sin \left( q_{\text{TF}}x\right) \left[ \frac{\pi }{2}-\text{Si}\left(
q_{\text{TF}}x\right) \right] \Bigr\},  \label{eq:V_TF}
\end{align}%
where $\text{Ci}$ and Si denote the cosine and sine integral functions, $%
\lambda $ is the line charge density and $q_{\text{TF}}=e^{2}k_{F}/(\pi
\varepsilon _{0}\varepsilon _{\text{r}}\hbar v_{F})$ is the TF
wavevector defined by the electron Fermi velocity $v_{F}=3ta/(2\hbar )$ and
the Fermi momentum $k_{F}=\sqrt{\pi n}$ ($a=0.142$~nm is the C-C distance).
The TF potential (\ref{eq:V_TF}) was used for semiclassical Boltzmann and
quantum-mechanical Kubo calculations of the conductivity of CVD graphene
\cite{Ferreira,Taras_Aires}. It is therefore important to find whether this
potential provides a reliable approximation for the numerically exact
quantum-mechanical self-consistent potential. A comparison between the TF
potential (\ref{eq:V_TF}) and the self-consistent potential calculated on
the basis of the lattice Hamiltonian Eq. (\ref{eq:Hamiltonian}) is shown in
Fig. \ref{fig:charge} (c),(d) for respectively (2,0) and (2,1) GBs. In this
comparison the electron density $n$ and the line charge density $\lambda $
in the TF potential (entering Eq. (\ref{eq:V_TF}) as phenomenological
parameters) have been extracted from our numerical calculations, see the
caption to Fig. \ref{fig:charge}. The overall agreement for the potential
height and width between the TF and the numerically exact potential is satisfactory.
However, the TF approach predicts a potential that is more narrow and decays
more rapidly in comparison to the exact one, especially for the (2,0) GB.
This can be related to a finite extent of the wavefunction disregarded in
the TF approach where a state residing at a GB is treated as a charged line
of a zero width. Note that because of the finite width of the ribbon the
long-distance behavior of the exact self-consistent potential is obscured by
edge effects. This makes it difficult to provide a quantitative comparison
of its long-distance asymptotic to that of the TF potential (\ref%
{eq:V_TF}) which decays as $V_{TF}(x)\propto \left( q_{\text{TF}}x\right)
^{-2}$.

It has been speculated in the literature that GBs can be used as a
one-dimensional quantum wire or a device component to carry the current in
bulk graphene \cite{Lahiri,Liwei,Peres}. Our findings suggest that these
predictions might be too optimistic. Indeed, GB states with a flat
dispersion (such as (2,0) states) remain metallic and pinned at $E_{F}$ even
at finite gate voltages when graphene is electrostatically doped by
electrons. However, the group velocity of such states is practically
zero, which makes these states hardly suitable for transport of current. On
the other hand, states such as (2,1), when populated at finite gate voltages,
move below $E_{F}$ thus losing their ability to carry current.

\begin{figure}[th]
\includegraphics[width=1\columnwidth]{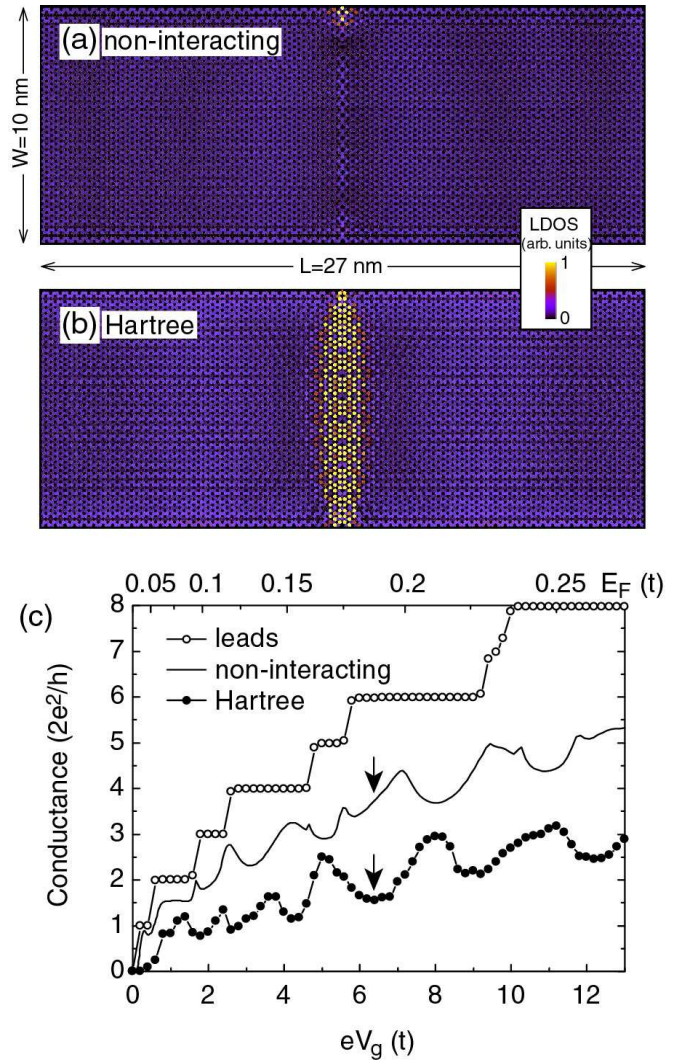}
\caption{(color online) (c) The conductance of a graphene nanoribbon with
the (2,0) GB. The LDOS calculated in (a) the noninteracting model and (b) the Hartree
model at $E_{F}=6.4t$ (marked by arrows in (c)). Ribbon's width $W=10$ nm; length $L=27$ nm, which
corresponds to 41 carbons in the transverse direction and 63 unit cells in
the longitudinal direction. }
\label{fig:cond}
\end{figure}

To study the effect of the GB on electron transport we consider the (2,0) GB
embedded in an armchair ribbon, see Fig. \ref{fig:cond}. The conductance of
an ideal ribbon shows quantized steps corresponding to the opening of new
transverse subbands, Fig. \ref{fig:cond}(c). The conductance of a ribbon with
the GB calculated for non-interacting electrons shows an overall drop of $%
\sim 30\%$ in comparison to the ideal case, and it exhibits an oscillating
behavior resulting from electron interference within the GB. Accounting for
electron interaction results in a further drop of the conductance ($\sim
\frac{2}{3}$ in comparison to the non-interacting case). The smaller conductance of
the GB for interacting electrons in comparison to non-interacting ones is
due to enhanced backscattering from the electrostatic potential at the GB
caused by the electrons accumulated there, see the LDOS in Figs. \ref{fig:cond}%
(a),(b). It is interesting to note that the Hartree approach predicts a
somewhat larger transport gap in comparison to the noninteracting case.

\begin{figure}[th]
\includegraphics[width=1\columnwidth]{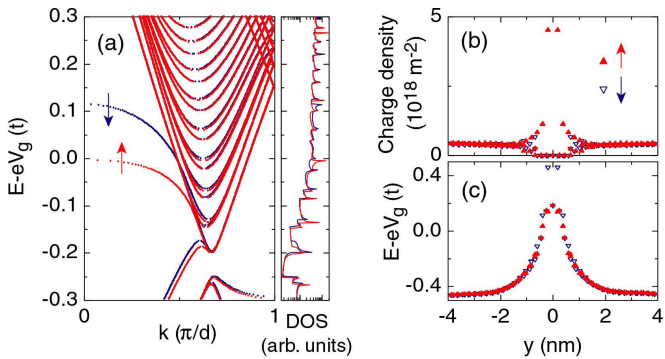}
\caption{(color online) (a) The band structure, (b) the charge density and
(c) the self-consistent potential calculated in the Hubbard-Hartree approach
for the (2,0) GB at $eV_{g}=5t$. }
\label{fig:spin}
\end{figure}

Let us finally explore spin polarization in the GBs. It has been shown
before that the (2,0) GB in neutral graphene is spin polarized with the
ferromagnetic ground state \cite{Botello,Alexandre,Okada,Kou}. Our
calculations based on the Hamiltonian Eq. (\ref{eq:Hubbard}) show that this state remains fully spin-polarized in
electrostatically doped graphene at a finite $V_{g}$ with the electron
population being completely dominated by species of the same spin, see Fig. %
\ref{fig:spin} for the spin resolved band structure, the electron density and
the potential of the (2,0) GB at $eV_{g}=5t$, (c.f. with the spin-degenerate
band structure of the same structure in Fig. \ref{fig:band}(b)). Due to the
metallic character of the flat state (2,0) which is partially filled at $%
E_{F}$, the electron density in this state can be easily redistributed.
Hence, the spin-up and spin-down states can have different densities and
thus experience different interaction due to the Hubbard term. As a result,
the spin-down state is pushed up by the Hubbard interaction above $E_{F}$
and gets depopulated while the spin-up state remains populated and pinned to
$E_{F}$. This is similar to the spin polarization of compressible strips in
quantum wires and graphene nanoribbons in a high magnetic field \cite%
{Marcus,Ihnatsenka2012}.

For the case of the (2,1) GB, the corresponding state is fully
occupied even for small applied $V_{g}$, and therefore the spin-up and
spin-down electron densities are the same. As a result, the potential felt
by different spin species is the same and the spin polarization for the
(2,1) state is completely suppressed.

\section{Conclusions}
We demonstrated that electronic, transport, and spin properties
of GBs are strongly modified in electrostatically doped graphene at finite electron densities in comparison to a
conventional noninteracting electron picture. Our calculations of the band structure and the conductivity were based on the self-consistent Green's function technique  where electron interactions were included by the Hartree potential (for spinless electrons) and by the Hartree and Hubbard potentials (for the spin-resolved case). Our main findings can be summarized as follows.

(1) We demonstrated that the character of charge accumulation is different for different GBs. In particular,  the charge density at the (2,0) GB grows linearly with an increase of the gate voltage $V_g$. In contrast,
the charge density at the (2.1) GB stays practically constant,
showing only a slow increase following the increase
of the overall electron density in the ribbon as $V_g$ grows. We analyzed in detail the band structure and related the above difference in the charge accumulation to the different characters of the band dispersions and the DOS of the states residing at the (2,0) and (2,1) GBs.

(2) We calculated the numerically exact self-consistent potential using Hamiltonian Eq. (\ref{eq:Hamiltonian}) and showed that this potential can be satisfactory approximated by the analytical expression, Eq. (\ref{eq:V_TF}), obtained within the semiclassical TF approximation.

(3) We studied the effect of the GB on electron transport by considering the (2,0) GB embedded in an armchair ribbon. We demonstrated that accounting for electron interaction results in a drop of the conductance ($\sim 2/3$ as compared to the non-interacting case). We relate this to
the enhanced backscattering from the electrostatic potential at the GB caused by the electrons accumulated there.

(4) In contrast to earlier speculations in the literature that GBs can
be used as one-dimensional quantum wires to carry the current in bulk graphene, our findings suggest that these predictions might be
over-optimistic. Even though the (2,0) GB state retains its metallic character even at finite electron densities, its group velocity is practically zero, which makes this state hardly suitable for the transport of current. As far as the (2,1) GB state is concerned, when populated at finite gate voltages it moves below $E_F$ thus losing its ability to carry current.

(5) For the spin-polarized case we found that in electrostatically doped graphene at a finite $V_g$ the state residing at the (2,0) GB  gets fully spin-polarized with the electron population being completely dominated by species of the same spin. In contrast, (2,1) GB states remain spin-degenerate. This difference is traced to different characters of the band dispersions of these states.

\begin{acknowledgments}
The authors gratefully acknowledge financial support from the Swedish
Institute and thank George Kirczenow and Aires Ferreira for discussions.
\end{acknowledgments}

\appendix

\section{Geometry relaxation and calculations of the hopping integrals}

\begin{figure*}[ht]
\includegraphics[width=0.9\textwidth,keepaspectratio]{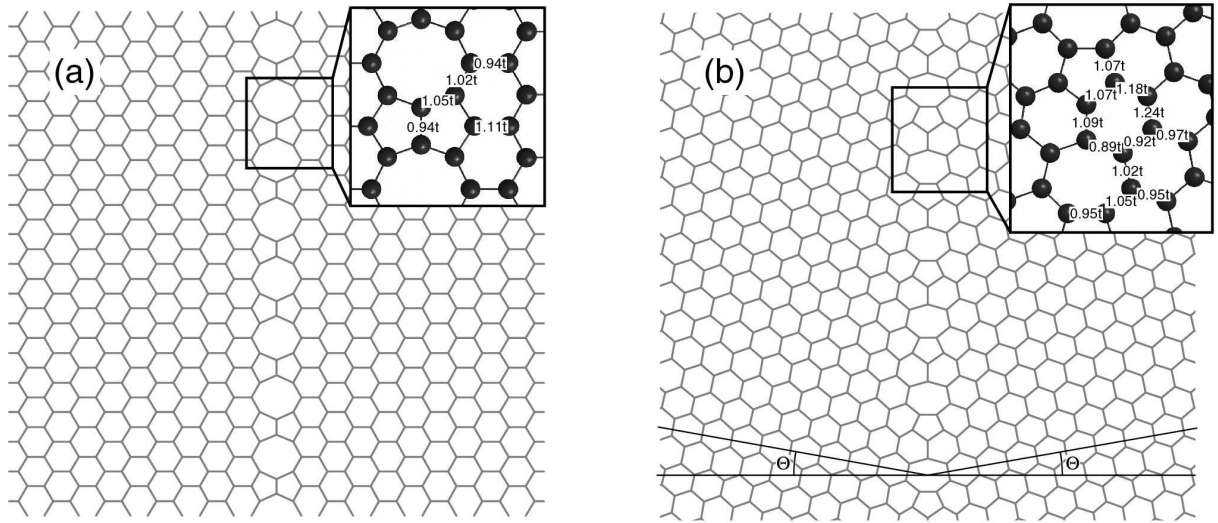}
\caption{ Relaxed atomic geometries with representative hopping
energies in units of $t$ for (2,0) and (2,1) GBs ((a) and (b) respectively). All geometries are in-plane. Grain boundaries separate two crystalline domains rotated
by different tilt angles $\theta = 0$ and $10.9 ^{\circ}$, (a) and
(b), respectively.}
\label{fig:SI:1}
\end{figure*}


We performed {\em ab initio} geometry relaxations based on the
density functional theory for the grain boundary defects using the
Gaussian 09 software package.\cite{Gaussian} The relaxed geometries
calculated in this way are expected to be accurate since the density functional
theory has been well optimized for carrying out the accurate ground
state total energy calculations.\cite{Kirczenow} The structures studied
were graphene flakes of about a hundred of carbon atoms passivated
at the edges with hydrogen with the grain boundaries being extended
across the flake. The atoms of the grain boundary as well as the nearest
carbon atoms were allowed to relax freely, while the other carbon
atoms were allowed to move normal to the defect line; the whole geometry
was kept planar. The relaxed structures obtained in this way are shown
in the insets in Fig. \ref{fig:SI:1}.

The tight binding model Hamiltonian, Eq. (1), includes
modified hopping energies between the graphene carbon atoms calculated
from the relevant matrix elements within the extended Hückel model.
The extended Hückel theory is formulated in terms of small basis sets
of Slater-type atomic orbitals $\left\{ \left|\phi_{i}\right\rangle \right\} $,
their overlaps $S_{ij}=\left\langle \phi_{i}|\phi_{j}\right\rangle $,
and the Hamiltonian matrix $\mathcal{H}_{ij}=\left\langle \phi_{i}\left|\mathcal{H}\right|\phi_{j}\right\rangle $.
The diagonal Hamiltonian elements $\mathcal{H}_{ii}=\mathcal{E}_{i}$
are chosen to be the experimentally determined atomic orbital ionization
energies $\mathcal{E}_{i}$. In the present work the nondiagonal elements
are approximated as in Ref. \onlinecite{yaehmop} by $\mathcal{H}_{ij}=(1.75+\Delta_{ij}^{2}-0.75\Delta_{ij}^{4})S_{ij}\left(\mathcal{E}_{i}+\mathcal{E}_{j}\right)/2$,
where $\Delta_{ij}=(\mathcal{E}_{i}-\mathcal{E}_{j})/(\mathcal{E}_{i}+\mathcal{E}_{j})$,
a form chosen to reproduce experimental molecular electronic structure
data \cite{yaehmop}. In the standard tight binding Hamiltonian of
pristine graphene, the energy scale is chosen such that the carbon
$2p_{z}$ orbital energy is zero whereas in the extended Hückel theory\cite{yaehmop}
the carbon $2p_{z}$ orbital energy is the ionization energy $\mathcal{E}_{\mathrm{C}_{p_{z}}}=-11.4$
eV. Accordingly, for consistency, in the extended Hückel Hamiltonian
matrix we make the replacement $\mathcal{H}_{ii}\rightarrow\mathcal{H}_{ii}-\mathcal{E}_{\mathrm{C}_{p_{z}}}$.
Because the extended Hückel basis states on different atoms are not,
in general, mutually orthogonal, the non-diagonal extended Hückel Hamiltonian
matrix elements are then also adjusted according to
\begin{equation}
\mathcal{H}_{ij}\rightarrow\mathcal{H}_{ij}-S_{ij}\mathcal{E}_{\mathrm{C}_{p_{z}}}.\label{eq:shift}
\end{equation}
Finally we extract the nearest-neighbor Hamiltonian matrix elements
$t_{ij}$ from $\mathcal{H}_{ii}$ and incorporate them into the standard
$\pi-$band Hamiltonian, Eq. (1), used for the self-consistent
calculations. For consistency with the standard tight-binding model
of pristine graphene we scale the values of $t_{ij}$ obtained from
the extended Hückel model as described in Ref. \onlinecite{adsorbate}.
The inserts in Fig. \ref{fig:SI:1} show the extracted hopping energies
$t_{ij}$.


\begin{thebibliography}{99}
\bibitem{Kim} K. S. Kim, Y. Zhao, H. Jang, S. Y. Lee, J. M. Kim, K. S. Kim ,
J. H. Ahn, P. Kim, J. Y. Choi, and B. H. Hong, Nature \textbf{457}, 706
(2009).

\bibitem{Li} X. Li, W. Cai, J. An, S. Kim, J. Nah, D. Yang, R. Piner, A.
Velamakanni, I. Jung, E. Tutuc, S. K. Banerjee, L. Colombo, R. S. Ruoff,
Science \textbf{324}, 1312 (2009).

\bibitem{Huang} P. Y. Huang, C. S. Ruiz-Vargas, A. M. van der Zande, W. S.
Whitney, M. P. Levendorf, J. W. Kevek, S. Garg, J. S. Alden, C. J. Hustedt,
Y. Zhu, J. Park, P. L. McEuen, and D. A. Muller, Nature \textbf{469}, 389
(2010).

\bibitem{KimCrommie} K. Kim, Z. Lee, W. Regan, C. Kisielowski, M. F.
Crommie, and A. Zettl, ACS NANO \textbf{5}, 2142 (2011).

\bibitem{Terrones} for a review see e.g. H. Terrones, R. Lv, M. Terrones and
M. S. Dresselhaus, Rep. Prog. Phys. \textbf{75}, 062501 (2012).

\bibitem{Lahiri} J. Lahiri, Y. Lin, P. Bozkurt, I. I. Oleynik, and M.
Batzill, Nature Nano. \textbf{5}, 326 (2010).

\bibitem{YazyevNature} O. V. Yazyev and S. G. Louie, Nat. Mater. \textbf{9},
806 (2010).

\bibitem{Yu} Q. Yu, L.A. Jauregui, W. Wu, R. Colby, J. Tian, Z. Su, H. Cao,
Z. Liu, D. Pandey, D. Wei, T.F. Chung, P. Peng, N.P. Guisinger, E.A. Stach,
J. Bao, S.-S. Pei, Y.P. Chen, Nat. Mater \textbf{10}, 443 (2011).

\bibitem{Ahmad} M. Ahmad, H. An, Y. S. Kim, J. H. Lee, J. Jung, S.-H. Chun
and Y. Seo, Nanotechnology \textbf{23}, 285705 (2012)

\bibitem{Tapaszto} L. Tapasztó, P. Nemes-Incze, G. Dobrik, K. J. Yoo, Ch.
Hwang, and L. P Biró, Appl. Phys. Lett. \textbf{100}, 053114 (2012)

\bibitem{Tsen} A. W. Tsen, L. Brown, M. P. Levendorf, F. Ghahari, P. Y.
Huang, R. W. Havener, C. S. Ruiz-Vargas, D. A. Muller, P. Kim, J. Park,
Science \textbf{336} 1143 (2012).

\bibitem{Koepke} J. C. Koepke, J. D. Wood, D. Estrada, Z.-Y. Ong, K. T. He,
E. Pop, and J. W. Lyding, ACS NANO \textbf{7}, 75 (2013).

\bibitem{Ferreira} A. Ferreira, X. Xu, C.-L. Tan, S.-K. Bae, N. M. R. Peres,
B.-H. Hong, B. Ozyilmaz, and A. H. Castro Neto, EPL, \textbf{94}, 28003
(2011).

\bibitem{Peres} J. N. B. Rodrigues, N. M. R. Peres, and J. M. B. Lopes dos
Santos, Phys. Rev. B \textbf{86}, 214206 (2012).

\bibitem{Tuan2013} D. V. Tuan, J. Kotakoski, T. Louvet, F. Ortmann, J. C.
Meyer, and S. Roche, Nano Lett. \textbf{13}, 1730 (2013).

\bibitem{Taras_Aires} T. M. Radchenko, A. A. Shylau, I. V. Zozoulenko, A.
Ferreira, Phys. Rev. B \textbf{87}, 195448 (2013).

\bibitem{Botello} A. R. Botello-Méndez, E. Cruz-Silva, F. López-Urías, B. G.
Sumpter, V. Meunier, M. Terrones, and H. Terrones, ACS NANO \textbf{3}, 3606
(2009).

\bibitem{Alexandre} S. S. Alexandre, A. D. Lúcio, A. H. Castro Neto, and R.
W. Nunes, Nano Lett. \textbf{12}, 5097 (2012).

\bibitem{Bahamon} D. A. Bahamon, A. L. C. Pereira, and P. A. Schulz, Phys.
Rev. B \textbf{83}, 155436 (2011).

\bibitem{Song} J. Song, H. Liu, H. Jiang, Q.-f. Sun, and X. C. Xie, Phys.
Rev. B \textbf{86}, 085437 (2012).

\bibitem{Liwei} L. Jiang, G. Yu, W. Gao, Z. Liu, and Y. Zheng, Phys. Rev. B
\textbf{86}, 165433 (2012).

\bibitem{Kou} L. Kou, C. Tang, W. Guo, and C. Chen, ACS NANO \textbf{5},
1012 (2011).

\bibitem{Okada} S. Okada, K. Nakada, K. Kuwabara, K. Daigoku, and T. Kawai,
Phys. Rev. B \textbf{74}, 121412R (2006).

\bibitem{Semenoff} G.W. Semenoff, V. Semenoff, and Fei Zhou, Phys. Rev.
Lett. \textbf{101}, 087204 (2008),

\bibitem{Jung} J. Jung, Z. Qiao, Q. Niu, and A. H. MacDonald, Nano Letters,
Nano Lett. \textbf{11}, 3453 (2011).

\bibitem{Qiao} Z. Qiao, J. Jung, Q. Niu, and A. H. MacDonald, Nano Letters,
Nano Lett. \textbf{12}, 2936 (2012).

\bibitem{Shylau_Cap} A. A. Shylau, J. W. Klos, and I. V. Zozoulenko, Phys.
Rev. B \textbf{80}, 205402 (2009).

\bibitem{Gaussian} GAUSSIAN 09, revision A.02 (Gaussian Inc., Wallingford,
CT, 2009); The HSEh1PBE hybrid density functional and 6-311G(d) basis set
were used in the geometry relaxations carried out in the present study.

\bibitem{Ihnatsenka2012} S. Ihnatsenka and I. V. Zozoulenko, Phys. Rev. B
\textbf{86}, 155407 (2012).

\bibitem{conductance} S. Ihnatsenka \& G. Kirczenow, Phys. Rev. B \textbf{86}%
, 075448 (2012).

\bibitem{Wehling} T. O. Wehling, E. \c{S}a\c{s}io\u{g}lu, C. Friedrich, A.
I. Lichtenstein, M. I. Katsnelson and S. Blügel, Phys. Rev. Lett. \textbf{106%
}, 236805 (2011).

\bibitem{Yazyev} O. V. Yazyev, Phys. Rev. Lett. \textbf{101}, 037203 (2008).

\bibitem{Marcus} S. Ihnatsenka and I. V. Zozoulenko, Phys. Rev. B \textbf{78}%
, 035340 (2008).


\bibitem{Kirczenow} For a recent review see G. Kirczenow, \textit{Molecular
nanowires and their properties as electrical conductors}, The Oxford
Handbook of Nanoscience and Technology,Volume I: Basic Aspects, Chapter
4, edited by A. V. Narlikar and Y. Y. Fu, Oxford University Press,
U.K. (2010).

\bibitem{yaehmop}The version of extended Hückel theory that we use
here is that of J. H. Ammeter, H.-B. Bürgi, J. C. Thibeault, and R.
Hoffman, J. Am. Chem. Soc. \textbf{100}, 3686 (1978) as implemented
in the YAEHMOP numerical package by G. A. Landrum and W. V. Glassey
(Source-Forge, Fremont, California, 2001).

\bibitem{adsorbate} S. Ihnatsenka and G. Kirczenow, Phys. Rev. B
\textbf{83}, 245442 (2011).
\end{thebibliography}
\end{document}